\documentclass[conference]{IEEEtran}
\IEEEoverridecommandlockouts
\usepackage{cite}
\usepackage{amsmath,amssymb,amsfonts}
\usepackage{graphicx}
\usepackage{textcomp}
\usepackage{xcolor}
\def\BibTeX{{\rm B\kern-.05em{\sc i\kern-.025em b}\kern-.08em
    T\kern-.1667em\lower.7ex\hbox{E}\kern-.125emX}}

\newcommand{\name}{LACE\xspace}

\usepackage[font=small,labelfont=bf]{caption}
\usepackage{subcaption}
\usepackage[normalem]{ulem}
\usepackage{amsmath,amssymb,amsfonts}
\usepackage{graphicx}
\usepackage{textcomp}
\usepackage{xcolor}
\usepackage{multirow}
\usepackage{booktabs}
\usepackage{hyperref}
\usepackage{tikz}
\usepackage{comment}
\usepackage{array}
\usepackage{algorithm}
\usepackage{algpseudocode}
\usepackage{threeparttable}
\usepackage{amssymb}
\usepackage{pifont}%

\usepackage[]{hyperref}
\usepackage{hyperxmp} 
\usepackage{enumitem}
\usepackage[textsize=footnotesize,textwidth=1.3\linewidth]{todonotes}

\usepackage{booktabs}
\usepackage{mathtools}
\usepackage{amsmath}
\usepackage{multirow} 
\usepackage{makecell}
\usepackage{times}
\usepackage{epsfig}
\usepackage{subcaption}
\usepackage{fancyhdr}
\usepackage[normalem]{ulem}
\usepackage{soul}
\usepackage{dblfloatfix}
\usepackage{todonotes}
\usepackage{enumitem}
\usepackage{booktabs}
\usepackage{etoolbox}
\usepackage{lipsum}
\usepackage{wrapfig}
\usepackage{algorithm}
\usepackage{algpseudocode}
\usepackage{colortbl}
\usepackage[table,xcdraw]{xcolor}
\usepackage{listings}
\usepackage{placeins} 

\usepackage{pifont}
\usepackage{xcolor}
  
\usepackage{xspace}
\usepackage{tabularx}

\newcommand{\inlineenum}[1]{%
    \ifcase#1%
        \or\ding{172}%
        \or\ding{173}%
        \or\ding{174}%
        \or\ding{175}%
        \or\ding{176}%
        \or\ding{177}%
        \or\ding{178}%
        \or\ding{179}%
        \or\ding{180}%
        \or\ding{181}%
        \else\ding{172}%
    \fi%
}

\begin{document}

\title{LACE: Large Language Model Aided Multi-Agent Framework for Agile RISC-V
Instruction Extension
}

\author{
\IEEEauthorblockN{Pingqing Zheng$^1$, Jiayin Qin$^1$, Fuqi Zhang$^1$, Zishen Wan$^2$, Shang Wu$^3$,\\ Yu (Kevin) Cao$^1$, Caiwen Ding$^1$, Yang (Katie) Zhao$^1$
}
\IEEEauthorblockA{\textit{
$^1$ University of Minnesota, Twin Cities,  Minneapolis, Minnesota, USA}
}
\IEEEauthorblockA{\textit{
$^2$ Columbia University, New York City, New York, USA}
}
\IEEEauthorblockA{\textit{
$^3$ Northwestern University, Evanston, Illinois, USA}
}
}


\maketitle

\begin{abstract}
Domain-specific Instruction Set Architecture eXtensions (ISAX) are widely adopted in the RISC-V ecosystem to accelerate emerging workloads, but implementing and validating ISAXes across different cores remains slow and fragmented. Existing frameworks still require per-core interface adaptation, and differential testing often breaks once either the microarchitecture or the ISAX changes. We present \name, an LLM-aided multi-agent workflow that translates natural-language ISAX intents into a compact two-level IR (operation-level and HDL task-level), performs retrieval-guided localized RTL edits over large repositories, and closes the loop with a compiler-agnostic \textit{riscv-formal} checking flow (assuming RVFI availability or instrumentation). Across four embedded RISC-V cores, \name raises pass@1 generation accuracy from near-zero to 72.8\% within our evaluation setup, while improving code localization and reducing integration rework.
The code of {\name} is available at https://github.com/UMN-ZhaoLab/LACE.
\end{abstract}

\begin{IEEEkeywords}
RISC-V, instruction set extension, large language model, multi-agent system
\end{IEEEkeywords}

\section{Introduction}
\label{sec:Introduction}

As emerging workloads push beyond the capabilities of conventional embedded and IoT cores, application-specific instruction-set processors (ASIPs) have become a practical solution. By extending base ISAs with domain-specific instructions and lightweight architectural modifications, ASIPs can improve workload efficiency while retaining toolchain compatibility and software fallback~\cite{hussein2024automating,302348,hepola2022openasip}. RISC-V’s open and modular ISA further makes it a natural foundation for ASIP development, but also encourages vendor- and project-specific extensions that can fragment software portability and ecosystem support. Although the RISC-V community is addressing this through standardized profiles (e.g., RV32A, RV32B), evaluating candidate instructions across diverse core implementations remains manual and time-consuming.

\begin{figure}
    \centering
    \includegraphics[width=\linewidth]{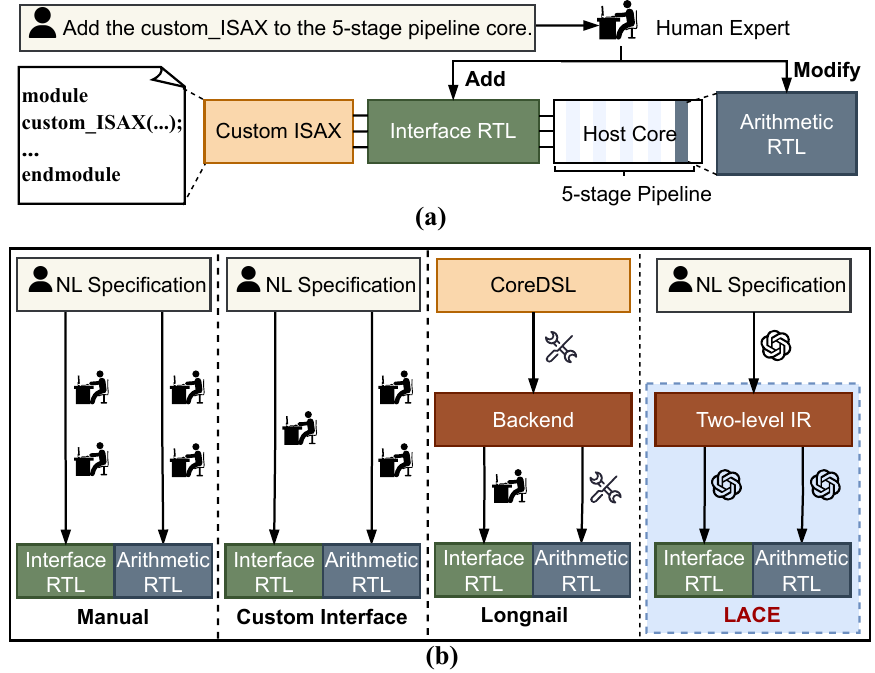}
    \vspace{-10pt}
    \caption{(a) An illustration of the conventional manual process for ISAX. (b) Comparison of our proposed {\name} with baseline works: one using a custom interface for updating RTL code~\cite{damian2022scaie} and another using a customized architecture description language (CoreDSL) for template-based RTL updates~\cite{oppermann2024longnail}.}
    \label{fig: compare}
\end{figure}

Existing automatic ISAX frameworks, such as SCAIE-V~\cite{damian2022scaie} and Longnail~\cite{oppermann2024longnail}, represent important progress toward scalable ISAX support (see Figure~\ref{fig: compare}(b)). However, adapting them to emerging workloads and unseen cores remains labor-intensive for two main reasons: \textbf{(1) Cross-disciplinary gaps in workload knowledge}: these frameworks rely on architecture descriptions authored by hardware designers, which can hinder reuse across domains; and \textbf{(2) Manual adaptation for unseen cores}: porting extensions across cores still demands substantial effort and often ties designers to specific tool chains or core families.



Recent advances in Large Language Models (LLMs) have demonstrated the ability to generate Hardware Description Language (HDL) code from high-level natural language descriptions, reducing the barrier for domain experts to contribute high-level knowledge in natural language. LLMs also have the potential to lower the effort of HDL generation, even for previously unseen hardware repositories.

However, directly integrating LLMs into the ISAX design workflow remains challenging for three main reasons: \textbf{(1) High abstraction gap between ISA and HDL}: ISA specifications describe high-level intent, while HDL requires detailed low-level modifications; \textbf{(2) Localization difficulty}: ISA extensions require precise edits in large hierarchical HDL repositories; and \textbf{(3) Limited verification coverage}: compiler-based customization restricts portability, while targeted tests cover only a narrow subset of ISA behaviors.

To bridge these gaps, we introduce \textbf{\name}, an LLM-aided unified framework for ISAX design and verification. \name uses multiple LLM agents for specification understanding and HDL code modification, together with HDL-related toolkits and the ReAct technique~\cite{yao2023react} to structure tool use. We present our contributions as follows:
\begin{itemize}
    \item We introduce \name, an LLM-aided framework that supports adapting RISC-V instruction set extensions across multiple microarchitectures, addressing the cross-core compatibility challenge that fragments the RISC-V ecosystem, unlike general RTL generators targeting isolated modules without architectural context.
    \item We design a two-level IR consisting of operation-level and HDL task-level representations that helps separate ISA semantics from microarchitecture-specific implementation details, enabling specification reuse across heterogeneous processor implementations.
    \item We develop an agentic RAG system for huge HDL code repository, improving the localization and modification of relevant code segments in large RISC-V CPU cores;
    \item We integrate riscv‑formal checking to enable compiler‑agnostic verification for newly generated instructions within our setup, strengthening the loop between ISAX design and functional validation;
    \item We demonstrate a working integration and verification flow and evaluate the practical benefits of our approach by extending five ISAXes into four open-source RISC-V cores and synthesizing the ISAX using TSMC 28nm technology.
\end{itemize}


\section{Background}
\label{sec: Background}

\subsection{RISC-V ISAX interface automation}

We identify three common ISAX integration styles: \inlineenum{1} in-pipeline ISAX, \inlineenum{2} custom coprocessor, and \inlineenum{3} dedicated accelerator. They trade off core coupling and communication overhead: in-pipeline ISAXes are suitable for lightweight latency-sensitive operations, while coprocessors and accelerators better serve heavier workloads through request-response or MMIO interfaces.

Existing commercial and open-source flows, including CodAL~\cite{codasip_codal}, Andes ACE~\cite{andes_ace_whitepaper}, EAI~\cite{e203_eai_spec}, CORE-V XIF~\cite{openhw_xif}, RoCC~\cite{yarp_rocc_intro}, PULPissimo~\cite{schiavone2018quentin}, and SCAIE-V~\cite{damian2022scaie}, cover different points in this design space. However, adapting these interfaces to a new microarchitecture still requires substantial manual effort to understand core-specific decode, control, datapath, and timing details. LACE targets this remaining per-core adaptation bottleneck by using LLM agents to infer and edit repository-specific integration patterns.

\subsection{LLM-aided hardware generation}
Recent studies~\cite{blocklove2023chip,chang2023chipgpt,lu2024rtllm,thakur2023autochip,pei2024betterv} have explored using LLMs to generate RTL code from natural language descriptions. Chip-Chat~\cite{blocklove2023chip} and Chip-GPT~\cite{chang2023chipgpt} employ LLMs with interactive human feedback, while RTLLM~\cite{lu2024rtllm} adopts a self-planning approach that outlines a high-level design plan before generation. Thakur et al.~\cite{thakur2023autochip} fine-tune large LLMs for HDL generation, and BetterV~\cite{pei2024betterv} improves dataset quality through data augmentation. However, none can yet generate repository-scale RTL systems such as RISC-V processors with high success rates. Three key gaps still hinder practical HDL-level processor design using LLMs:
\textbf{Gap 1: From natural language to HDL under strict semantic–implementation constraints.}
HDL data represents only about 1/800 of all programming languages~\cite{li2023starcoder}, and its abstractions differ fundamentally from software code~\cite{li2024agon}. As a result, LLMs struggle to align high-level semantics with detailed HDL behavior, especially in ISAX-related modifications.
\textbf{Gap 2: Between limited long-context reasoning and large hierarchical processors.}
LLM accuracy declines near context-window limits~\cite{liu2023lostmiddlelanguagemodels}. Even models with 128k-token windows (e.g., Qwen3-32B~\cite{yang2025qwen3}) cannot fully capture large designs such as cv32e40p~\cite{7864441}, which exceeds 243k tokens. Complex inter-module dependencies further challenge current LLMs’ hierarchical understanding.
\textbf{Gap 3: Between high generation error rates and reliable verification requirements.}
LLMs produce substantially more HDL errors than software code~\cite{pei2024betterv}. RTLFixer~\cite{tsai2024rtlfixer} reduces syntax issues through retrieval-augmented generation and ReAct prompting, but ensuring functional correctness and coverage in large systems remains an open problem.

\begin{figure*}[!t]
    \centering
    \includegraphics[width=\linewidth]{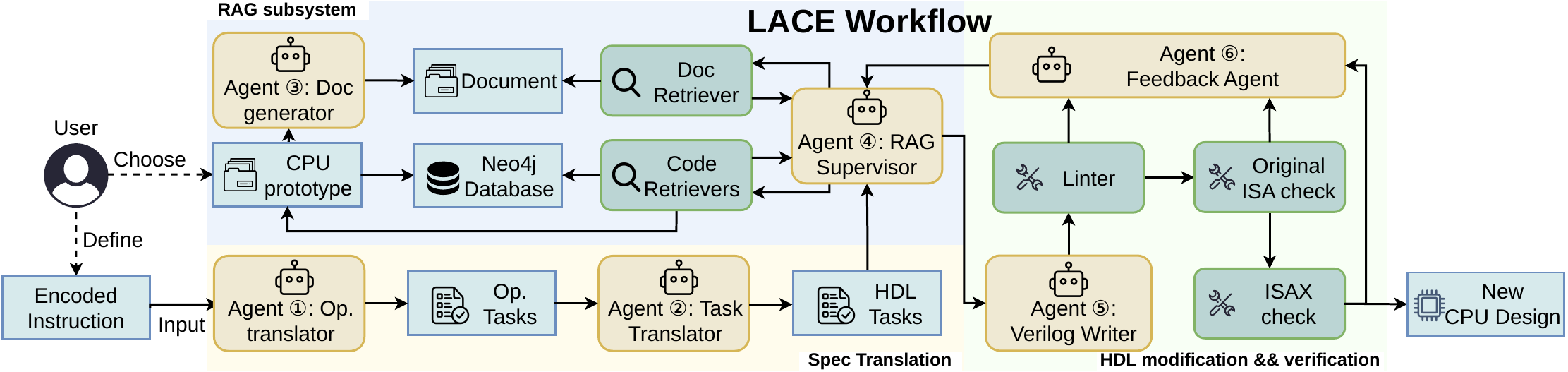}
    \vspace{-10pt}
    \caption{The overall \name workflow with specification translation, agentic RAG system and riscv-formal~\cite{riscv-formal} based verification.}
    \label{fig: overview}
\end{figure*}

\subsection{LLM-aided cross-hardware code optimization}
Recent work shows that LLMs can support cross-platform code optimization and migration, including semantic-preserving transformation~\cite{chen2023llmcompiler}, meta-prompting and auto-tuning for GEMM kernels on RISC-V/CUDA~\cite{qimeng2023}, OpenMP-to-CUDA translation with verification feedback~\cite{10.1007/978-3-031-85700-3_12}, and hardware-aware DSL generation for tensor accelerators~\cite{hong2024llmaidedcompilationtensoraccelerators}. LACE extends this direction from software/kernel translation to cross-microarchitecture RTL adaptation, where agents must modify processor-specific decode, control, datapath, interface timing, and verification logic for automated ISAX integration.

\section{LACE Workflow}
\label{sec: overview}

\subsection{Overview}

Figure~\ref{fig: overview} depicts a comprehensive workflow of~\name. Inspired by the collaborative division of labor in manual ISAX process, \name adopts a multi-agent architecture in which several LLM agents cooperate to automate ISAX integration. 
Specifically, \textbf{Agent \inlineenum{1} Operator Translator} first converts natural-language ISAX specifications into high-level architectural operations (Ops), which represent fundamental behaviors such as register accesses, pipeline control actions, and arithmetic primitives. These operations provide an abstraction layer that bridges instruction semantics and hardware implementation. \textbf{Agent \inlineenum{2} Task Translator} then maps these abstract operations into concrete HDL modification tasks according to the characteristics of the target processor microarchitecture. Meanwhile, \textbf{Agent \inlineenum{3} Document Generator} automatically generates and maintains architectural documentation of the target CPU, helping LLM agents better understand module organization and design constraints during the code generation process. To support architecture-aware code generation, \textbf{Agent \inlineenum{4} RAG Supervisor} retrieves relevant RTL code segments from the CPU repository, providing contextual information for accurate modifications. Based on this retrieved context, \textbf{Agent \inlineenum{5} Verilog Writer} performs precise file-level edits to insert or update RTL logic within the existing processor design. Each HDL modification is first validated using Verilator~\cite{Snyder_Verilator}. Syntax errors and simulation results are then analyzed by \textbf{Agent~\inlineenum{6} Feedback Agent}, which provides diagnostic feedback to the writer agent for correction. During the verification process, \name first checks the original ISA behavior and then verifies the functionality of the newly introduced ISAX. This iterative loop of generation and verification enables fast and reliable integration of ISAX. \name uses LangChain~\cite{Chase_LangChain_2022} to leverage its mature orchestration and tool integration.
\begin{table}[!t]
    \renewcommand{\arraystretch}{1.2}
    \caption{Predefined Interface Operators of ISAX Tasks.}
    
    \centering
    \small
    \begin{tabular}{@{}>{\raggedright\arraybackslash}p{0.3\columnwidth} 
                    >{\raggedright\arraybackslash}p{0.7\columnwidth}@{}}
    \toprule
    \textbf{Operations} & \textbf{Descriptions} \\ 
    \midrule
    RdInstr() & Read the instruction \\ 
    RdRS1/2 & Read register value based on rs1/2 \\ 
    CreateCustReg(num) & Create custom register file with `num` entries \\ 
    Rd/WrCustReg & Read or write value from custom register \\ 
    Rd/WrPC & Read or write current program counter value \\ 
    Rd/WrMem & Read from or write to memory \\ 
    WrRD & Write to register specified by rd field \\ 
    \bottomrule
    \end{tabular}
    \label{tab: predefined interface operations}
    \vspace{-10pt}
\end{table}

\begin{figure}[!b]
\vspace{-10pt}
    \centering
    \includegraphics[width=\linewidth]{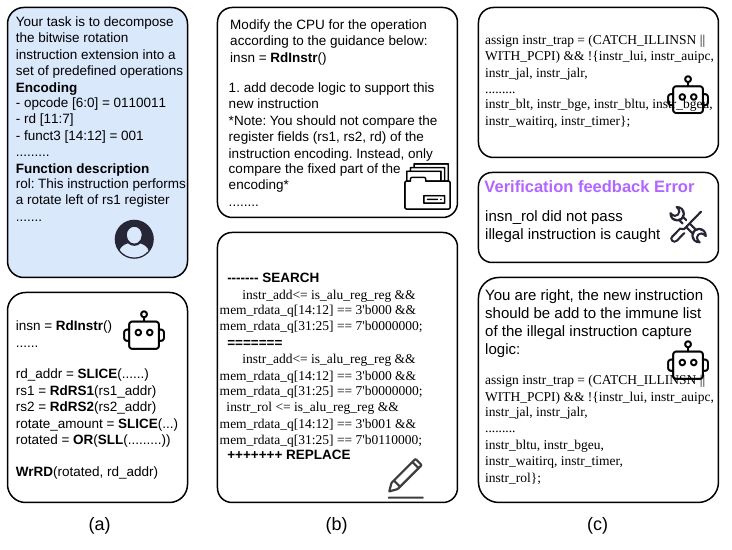}
    \vspace{-17pt}
    \caption{Prompt Examples of \name. From left to right are examples of operation generation, HDL modification, and debugging.}
    \label{fig: prompt example}
\end{figure}


Figure~\ref{fig: prompt example} illustrates an example of this workflow using a bit-rotation instruction, among which the first two agents address the abstraction gap between ISA specifications and HDL implementations. \textbf{Agent~\inlineenum{1}} translates natural-language ISAX specifications into architectural operations, while \textbf{Agent~\inlineenum{2}} further maps these operations into architecture-aware HDL modification tasks. 
By lifting the input specification to the natural-language level, this design enables workload experts to directly participate in the ISAX development process. 
Sections~\ref{subsec: spec trans} and~\ref{subsec: hdl modify workflow} further detail the specification translation and HDL adaptation workflows.

\subsection{Agent~\inlineenum{1}: Specification to Operator Translation}
\label{subsec: spec trans}

Existing ISAX automation frameworks, such as SCAIE-V~\cite{damian2022scaie} and Longnail~\cite{oppermann2024longnail}, rely on DSL-style specifications that still require hardware expertise. To allow ISAX design to start from natural-language intent, \name introduces \textbf{Agent~\inlineenum{1} Operator Translator}, which maps ISA descriptions to a compact operation-level IR extending prior operator-based representations~\cite{damian2022scaie,li2024agon}. Instead of generating RTL directly, the IR decomposes an instruction into human-readable and tool-actionable operators, as shown in Figure~\ref{fig: prompt example}(a--c) for the bit-rotation example.

The operators are divided into interface-related and arithmetic-related categories. Interface operators describe interactions with architectural state and pipeline control, such as reading registers, accessing the program counter, or writing back to \texttt{rd} (Table~\ref{tab: predefined interface operations}); these operators guide localized RTL edits such as adding decode paths, register ports, or handshake signals. Arithmetic operators capture functional semantics using composable arithmetic, bitwise, shift, and conditional primitives, while complex functions such as \textit{sincos} can be supported through custom hardware interfaces. This separation lets Agent~\inlineenum{1} preserve instruction semantics while leaving core-specific integration details to the downstream task translator.

\subsection{Agent~\inlineenum{2}: Operator to Task Translation}
\label{subsec: hdl modify workflow}

Although \name maps natural language ISAX specifications into predefined operations, large language models still struggle to perform accurate HDL edits. To guide them, \name introduces rigorously defined meta-prompts that encapsulate common implementation patterns across microarchitectures (e.g., state-machine and pipelined cores). Each meta-prompt contains an architecture-agnostic description, architecture-specific hints, and an instantiation procedure.

\textbf{Architecture-agnostic description.} This component provides general ISAX integration strategies of ISAX integration strategies with expert-designed templates that reduce reasoning complexity and improve LLM accuracy.

\textbf{Architecture-specific hints.} Hints refine the general strategies for specific cores by describing properties such as pipeline depth and control paths. For instance, updating the program counter in a single-cycle or state-machine CPU only requires reading the next instruction, whereas a pipelined CPU must also flush subsequent stages. \name maintains a modular hint repository of such hints that can be flexibly combined to form context-aware prompts.

For decoupled or long-latency ISAXes, the task translator emits additional control tasks such as request/response handshakes, stalls, flushes, and writeback synchronization according to the core-specific hints.
\section{Agent-guided HDL RAG}
\label{sec: rag}

Since ISAX integration requires repeated edits to complex CPU code while guidance is expressed in natural language, accurate localization is essential. \name reuses previously published graph-based HDL RAG framework \cite{zheng2025hdlxgraphbridginglargelanguage} as the code-navigation backend and adapts it to the multi-agent ISAX workflow. Figure \ref{fig3: rag supervisor} shows the resulting architecture.

\subsection{Agent \inlineenum{3}: Database construction and tool design}
\label{sec: rag tools}

We briefly summarize the components used in \name. Agent \inlineenum{3} extracts essential pipeline information from the CPU prototype into a document for further architecture-aware guided HDL modification. Meanwhile, following \cite{zheng2025hdlxgraphbridginglargelanguage}, we represent the codebase as a graph-structured repository that provides structural context beyond plain text. Concretely, modules and always-blocks are indexed as nodes, connected by hierarchy and dataflow edges, and retrieved through embedding-based matching with guided traversal. Within \name, this backend exposes three tool categories:
\textbf{First-step localization:} embedding-based retrieval maps a natural-language query to a small set of candidate graph nodes.
\textbf{Guided graph traversal:} controlled hierarchy and dataflow traversal expands these candidates into a focused structural view.
\textbf{Free code exploration:} when the predefined interfaces are insufficient, agents can issue Cypher queries or textual exploration. The tool interface returns (a) a compact structural summary and (b) provenance pointers such as file paths and line ranges, allowing the writer agent to request the corresponding source slice.

\subsection{Agent \inlineenum{4} for Codebase Exploration}
\label{sec: agent orchestration}
\begin{figure}[t]
     \centering
\includegraphics[width=1.0\linewidth]{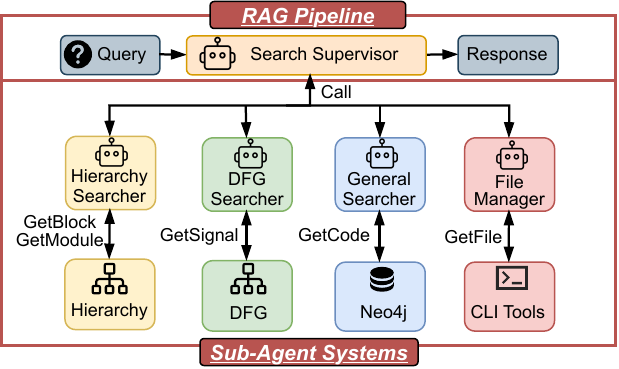} 
     \vspace{-15pt}
    \caption{Overview of the agentic HDL RAG system.}
    \label{fig3: rag supervisor}
\end{figure}

Recent studies demonstrate that supervisor-structured multi-agent systems achieve strong performance in complex reasoning and exploration tasks \cite{Orwall_Moatless_Tools_2024,wang2025openhands}. \name adopts a similar architecture for HDL code navigation, as illustrated in Figure~\ref{fig3: rag supervisor}. The supervisor coordinates specialized search agents to decompose, parallelize, and refine localization tasks.

\textbf{Search supervisor.} The supervisor manages query decomposition, task scheduling, and final synthesis. It decides whether a query primarily requires hierarchy navigation, dataflow traversal, or direct textual inspection, and then dispatches subtasks accordingly.

\textbf{Search sub-agents.}
Each sub-agent operates with a limited tool set and a bounded context window, then returns a concise summary with confidence and provenance information. If evidence is insufficient, the sub-agent reports failure in a structured form so the supervisor can re-plan. This keeps the retrieval process lightweight while still preserving traceability for downstream HDL modification.

\section{Verification Strategy}
\label{sec: verfication}

\subsection{Tool selection}

\begin{table}
\centering
\caption{ISAX benchmark used in the evaluation.}
\vspace{-5pt}
\begin{tabularx}{\columnwidth}{>{\raggedright\arraybackslash}p{1.8cm} >{\raggedright\arraybackslash}X >{\raggedright\arraybackslash}X}
\hline
\textbf{ISAX} & \textbf{Description} & \textbf{Category} \\ 
\hline
rol & Bitwise rotation \newline left & Simple combinational operation \\ \hline 
ijmp & Read next PC \newline from main memory & Custom control-flow \\ \hline
sbox & Lookup from \newline AES S-Box & Custom and long-latency \\ \hline
load\_mul & Load multiple & Main memory access \\ \hline
sincos & decoupled sine and cosine computation & Decoupled complex arithmetic \\
\hline
\end{tabularx}
\label{tab: instruction table}
\end{table}

Recent RISC-V verification efforts primarily rely on differential testing \cite{micro2022xiangshan,encore}, which compares the CPU state against a golden model cycle by cycle. Although effective for detecting discrepancies, applying this approach to LLM aided ISAX verification faces several challenges.

\textbf{Golden Model Correctness.} Differential testing assumes the golden model is bug-free, an unrealistic guarantee when LLMs modify or extend the architecture. LLM-generated models may contain subtle errors, requiring separate verification of the golden model itself and still leaving residual risk.

\textbf{Compiler Adaptation.} 
Differential testing depends on compiler support to generate test programs or issue random instructions. Adapting compilers via LLMs is nontrivial \cite{chen2023llmcompiler}, while manually embedding targeted instructions directly in test programs limits coverage, creating verification gaps.
To address these limitations, \name employs \textit{riscv-formal}~\cite{riscv-formal}, which verifies instruction behavior via RVFI-based assertions using Yosys \texttt{sby}. This eliminates reliance on golden-model correctness and avoids requiring a customized compiler frontend for each new instruction.

Nonetheless, \textit{riscv-formal} has its own constraints:
\inlineenum{1} Scalability. Formal verification is computationally expensive and may not scale to high-performance processors.
\inlineenum{2} RVFI Integration. Most existing cores lack native RVFI support, and inserting RVFI hooks automatically across diverse designs remains challenging. In this work, we assume RVFI is either already available or can be added with engineering effort; fully automatic RVFI insertion is outside the current scope.

\subsection{Agent \inlineenum{6} Stepwise Verification}
Because verifying ISAX extensions before completing interface integration is risky, where each modification may disrupt core functionality, \name adopts a stepwise approach. We first run \textit{riscv-formal} on the unmodified baseline core to establish a regression-safe reference. We then integrate the ISAX and re-run \textit{riscv-formal} with the extended decode and semantics, so each iteration checks both (i) preservation of baseline ISA behavior and (ii) the new instruction property set. This staged verification helps preserve correctness while reducing regression risks during iterative generation.

\section{Evaluation}
\label{sec: evaluation}



\subsection{Experiment Setup}

\textbf{Instruction benchmark.} 
As no standard ISAX benchmark exists, we define a custom set (Table~\ref{tab: instruction table}) covering three categories: (1) \textbf{Main memory access} (\textit{indirect jump}, \textit{load multiple}), (2) \textbf{Simple combinational} (\textit{bitwise rotation}), and (3) \textbf{Custom long-latency} (\textit{sbox},  \textit{sincos}) instructions. This set exercises memory interaction, lightweight datapath extension, and multi-cycle execution to evaluate \name comprehensively.

\begin{table*}
\centering
\caption{Generation accuracy comparison with non-agentic generation with GPT-4o.}

\begin{tabular}{lcccccccc}
\hline
 & \multicolumn{2}{c}{PicoRV32} & \multicolumn{2}{c}{e203\_hbird} & \multicolumn{2}{c}{ibex} & \multicolumn{2}{c}{cv32e40x} \\ 
\cline{2-9}
 & Non-agentic & \name & Non-agentic & \name & Non-agentic & \name & Non-agentic & \name \\ 
\hline
rol & 22.2\% & \cellcolor[HTML]{B6D7A8}100\% & 0 & \cellcolor[HTML]{B6D7A8}100\% & 11.1\% & \cellcolor[HTML]{B6D7A8}100\% & 11.1\% & \cellcolor[HTML]{B6D7A8}100\% \\ 
ijmp & 0 & \cellcolor[HTML]{B6D7A8}88.9\% & 0 & \cellcolor[HTML]{B6D7A8}55.6\% & 0 & \cellcolor[HTML]{B6D7A8}100\% & 0 & \cellcolor[HTML]{B6D7A8}66.7\% \\ 
sbox & 0 & \cellcolor[HTML]{B6D7A8}88.9\% & 0 & \cellcolor[HTML]{B6D7A8}77.8\% & 0 & \cellcolor[HTML]{B6D7A8}88.8\% & 0 & \cellcolor[HTML]{B6D7A8}55.6\% \\ 
load\_mul & 0 & \cellcolor[HTML]{B6D7A8}66.7\% & 0 & \cellcolor[HTML]{B6D7A8}44.4\% & 0 & \cellcolor[HTML]{B6D7A8}66.7\% & 0 & \cellcolor[HTML]{B6D7A8}55.6\% \\ 
sincos & 0 & \cellcolor[HTML]{B6D7A8}55.6\% & 0 & \cellcolor[HTML]{B6D7A8}55.6\% & 0 & \cellcolor[HTML]{B6D7A8}44.4\% & 0 & \cellcolor[HTML]{B6D7A8}44.4\% \\ 
\hline
\end{tabular}
\label{tab: pass rate}
\end{table*}

\textbf{Baseline cores.}
We evaluate a representative set of open-source embedded cores, PicoRV32 \cite{picorv32}, e203\_hbirdv2 \cite{nuclei_hbirdv2}, ibex \cite{lowrisc_ibex}, and cv32e40x \cite{7864441}, that collectively capture the architectural diversity of the RISC-V ecosystem. Among them, cv32e40x, ibex and e203\_hbirdv2 implement the four-stage pipeline, three-stage pipeline and two-stage pipeline respectively, while PicoRV32 is a non-pipelined core that relies on an FSM for instruction sequencing. The baseline core area and performance are reported in Table \ref{tab: PPA}.

\textbf{Configurations.} 
\name uses GPT-4o (temperature~0.8) with iterative syntax and functional checks via Verilator~\cite{Snyder_Verilator} and \textit{riscv-formal}. Unless otherwise stated, each (core, ISAX) pass@1 entry is computed over nine independent runs with the same 2-shot examples. A run is counted as successful only if it (i) compiles cleanly with Verilator and (ii) passes the \textit{riscv-formal} checks for the baseline core plus the added instruction. For each (core, ISAX), pass@1 is the fraction of nine independent complete LACE runs whose final RTL both compiles with Verilator and passes riscv-formal. Designs are synthesized in TSMC~28\,nm using Design Compiler to compare baseline and ISAX-extended cores.

\subsection{ISAX Generation with \name}

\textbf{Accuracy improvement.}
We compare the pass@1 accuracy of ISAX generation between \name and direct (non-agentic) LLM generation using identical target files, ISAX descriptions, 2-shot examples, and model settings. As shown in Table~\ref{tab: pass rate}, \name achieves an average pass rate of 72.8\%, while direct generation succeeds only for the simple \textit{bitwise rotation} instruction. This demonstrates that customized IRs, localized HDL edits, and verification feedback significantly improve reliability. Although success rates decline for more complex cores, all remain above 40\%, confirming that \name can autonomously generate most ISAXes without manual intervention.

\textbf{Performance and area overhead.} 
Table~\ref{tab: PPA} summarizes ASIC synthesis results. All frequencies are post-synthesis estimates under our stated constraints, rather than post-place-and-route results, and should therefore be interpreted only as relative comparisons within this setup. ISAX integration increases area by about 10\% and reduces frequency by less than 10\% in most cases, showing acceptable cost for added flexibility. Larger cores exhibit lower relative overhead, while complex multi-cycle instructions (\textit{sbox}, \textit{load\_mul}) introduce higher area and timing penalties.

\textbf{PPA overhead comparison with SCAIE-V \cite{damian2022scaie}.}
In order to compare the performance results with human expert's ISAX, we compare \name with SCAIE-V \cite{damian2022scaie} on a common core PicoRV32. Table~\ref{tab: PPA with SCAIE-V} demonstrates that with LLM's flexibility, \name can achieve a comparable performance with extension framework made by human expert.

\begin{table*}[t]
\centering
\caption{ASIC results for area and frequency overheads of ISAX when integrated into base cores. }

\begin{tabular}{lcccccccc}
\hline
 & \multicolumn{2}{c}{PicoRV32} & \multicolumn{2}{c}{e203\_hbird} & \multicolumn{2}{c}{ibex} & \multicolumn{2}{c}{cv32e40x} \\ 
\cline{2-9}
 & Area & Freq. & Area & Freq. & Area & Freq. & Area & Freq. \\ 
\hline
Base core & 5,662$\mu$m² & 3.70 GHz & 4,695$\mu$m² & 1.54 GHz & 6,591$\mu$m² & 1.75 GHz & 12,200$\mu$m² & 1.61 GHz \\ 
\hline
rol & \cellcolor[HTML]{B6D7A8}+2\% & \cellcolor[HTML]{B6D7A8}-2\% & \cellcolor[HTML]{B6D7A8}+2\% & \cellcolor[HTML]{B6D7A8}-0\% & \cellcolor[HTML]{B6D7A8}+1\% & -2\% & \cellcolor[HTML]{B6D7A8}+0\% & -1\% \\ 
ijmp & \cellcolor[HTML]{B6D7A8}+2\% & -3\% & +5\% & -4\% & +4\% & \cellcolor[HTML]{B6D7A8}-1\% & +9\% & \cellcolor[HTML]{B6D7A8}-0\% \\ 
sbox & +7\% & \cellcolor[HTML]{EA9999}-7\% & +10\% & \cellcolor[HTML]{EA9999}-8\% & +6\% & -3\% & +8\% & -1\% \\ 
load\_mul & +22\% & -6\% & +11\% & -5\% & +13\% & \cellcolor[HTML]{EA9999}-8\% & +10\% & \cellcolor[HTML]{EA9999}-7\% \\ 
sincos & \cellcolor[HTML]{EA9999}+69\% & -4\% & \cellcolor[HTML]{EA9999}+46\% & -3\% & \cellcolor[HTML]{EA9999}+22\% & -3\% & \cellcolor[HTML]{EA9999}+14\% & -3\% \\ 
\hline
\end{tabular}
\label{tab: PPA}
\end{table*}

\begin{table}[t]
\centering
\caption{Comparison of ASIC results for area and frequency overheads of ISAX with SCAIE-V~\cite{damian2022scaie} when integrated into PicoRV32.}

\begin{tabular}{lcccc}
\hline
 & \multicolumn{2}{c}{\name} & \multicolumn{2}{c}{SCAIE-V}  \\ 
\cline{2-5}
 & Area & Freq. & Area & Freq. \\ 
\hline
Base core & 5,662$\mu$m² & 3.70 GHz & 5,662$\mu$m² & 3.70 GHz  \\ 
\hline
rol & +2\% & -2\% & +2\% & -2\% \\ 
ijmp & \cellcolor[HTML]{B6D7A8}+2\% & -3\% & +7\% & -1\% \\ 
sbox & +7\% & \cellcolor[HTML]{EA9999}-7\% & +6\% & -2\% \\ 
load\_mul & +22\% & -6\% & +20\% & -4\% \\ 
sincos & +69\% & -4\% & +70\% & -4\% \\ 
\hline
\end{tabular}
\label{tab: PPA with SCAIE-V}
\end{table}

\subsection{Search Ability with Agentic RAG}

The agentic RAG system in LACE is capable of accurately locating and modifying complex HDL code in repository-level. To comprehensively evaluate its improvement in search capability on such codes, we construct a benchmark of 50 tests by collecting the given meta-prompts and their corresponding code segments to be modified. We adopt the widely used mean reciprocal rank ($\text{MRR} = \frac{1}{N} \sum_{i=1}^{N} \frac{1}{\text{rank}_i}$) in RAG as the primary evaluation metric, which measures whether the framework can return correct results among the top-ranked outputs.

\begin{figure}[t]
     \centering
\includegraphics[width=0.88\linewidth]{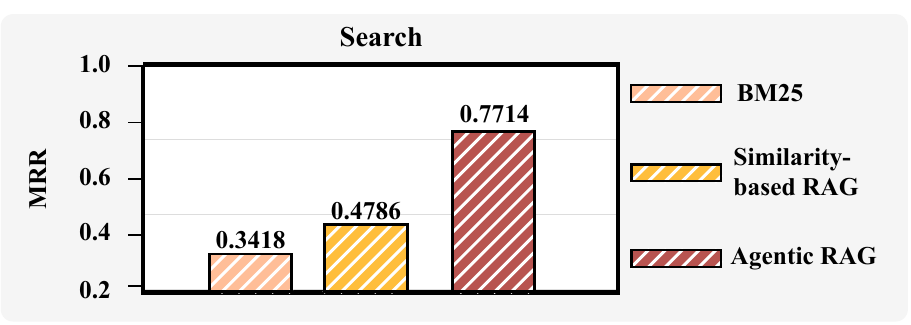} 
    \vspace{-4pt}
    \caption{Search Ability analysis with agentic RAG.}
    \label{fig :search ability}
\end{figure}

We compare {\name} against two commonly used similarity-based RAG methods, BM25 \cite{robertson2009probabilistic} and CodeT5+ embeddings \cite{wang-etal-2023-codet5}. As shown in Figure \ref{fig :search ability}, \name outperforms the baselines in block-level search, achieving an average MRR improvement of 29.28\% across all queries, demonstrating its potential for precise HDL search in complex repository-level codes.


\subsection{Convergence Efficiency with Multi-hop Verification}
\begin{figure}[t]
     \centering
\includegraphics[width=0.88\linewidth]{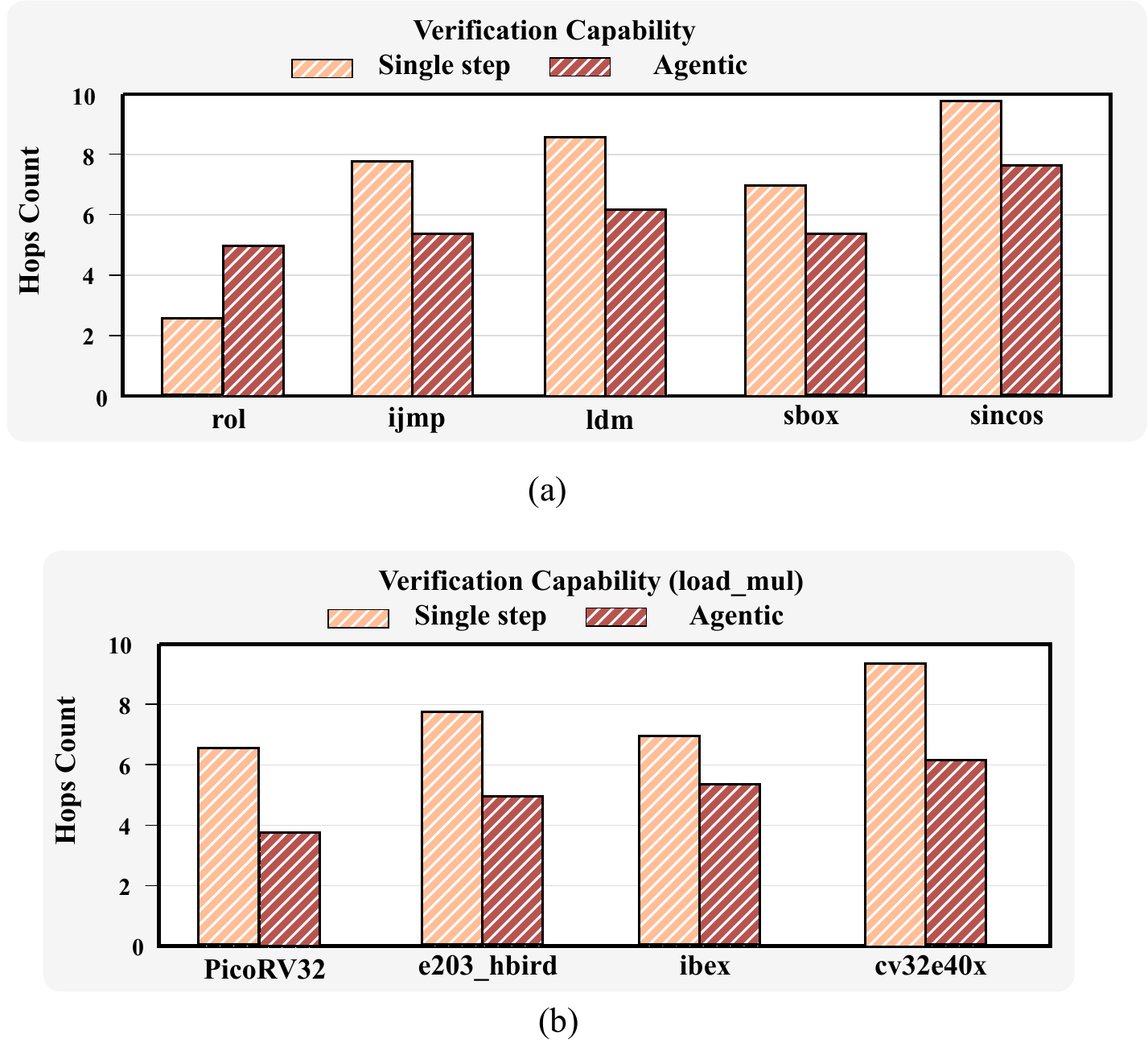} 
    \vspace{-4pt}
    \caption{Convergence efficiency with multi-hop verification with (a) 5 ISAXs on e203\_hbird and (b) load\_mul on 4 host cores.}
    \label{fig :veri hop}
\end{figure}

We assess how the multi-hop verification loop affects ISAX integration on the e203\_hbirdv2 core by comparing runs with and without incremental checking. As shown in Fig.~\ref{fig :veri hop}(a), although multi-hop verification adds minor overhead in simple designs, it greatly reduces total iterations for complex ones by preventing early errors from propagating. Figure~\ref{fig :veri hop}(b) further shows that for the \textit{load\_multiple} instruction, this approach consistently lowers verification hops across all evaluated cores.

Overall, LACE is more reliable for localized combinational extensions than for memory-coupled and long-latency ISAXes. Stepwise verification helps contain error propagation and reduce repair iterations. The remaining gap suggests that multi-cycle control and cross-module coordination require more specialized integration guidance.

\section{Conclusion and Future Work}
\label{sec:conclusion}

In this work, we presented \name, an LLM-aided multi-agent framework for agile RISC-V ISAX design and integration. LACE combines a reusable two-level IR, architecture-aware repository exploration, localized RTL editing, and formal-verification feedback to bridge natural-language ISAX specifications and processor-specific implementations.

Across four RISC-V cores and five instruction extensions, LACE achieves an average pass@1 of 72.8\%, substantially outperforming direct non-agentic generation. It also improves HDL localization and reduces verification iterations for complex extensions, while producing area and frequency overheads comparable to an expert-designed ISAX integration framework. Together, these results validate LACE as a reliable and efficient approach to cross-core ISAX integration.

Future work will extend \name to processors with more diverse and complex microarchitectures, including deeper pipelines, out-of-order execution, and more heterogeneous extension interfaces. We also plan to automate the processor-side RVFI instrumentation process, further reducing the manual effort required to enable formal verification. Additional directions include PPA-aware RTL optimization, automated compiler customization, and workload-level evaluation of candidate instruction extensions. These capabilities would allow \name to evolve from an ISAX implementation framework into a more comprehensive platform for automated instruction-set exploration, integration, and evaluation.

\section*{Acknowledgment}
    This work was partially supported by an Amazon Research Award (PIs: Profs. Yang (Katie) Zhao and Caiwen Ding).

\clearpage
\bibliographystyle{IEEEtran}
\bibliography{references,ref,ding}

\end{document}